\documentclass[twocolumn]{jpsj2}

\title{Thermodynamic Properties of the Heisenberg Antiferromagnet on a Railroad-Trestle Lattice with Asymmetric Leg Interactions}
\author{Shunsuke {\sc Takemura} and Yoshiyuki {\sc Fukumoto}}
\inst{Department of Physics, Faculty of Science and Technology, Tokyo University of Science, Noda, Chiba 278-8510}
\recdate{\hspace{3cm}}
\abst{
Using an approximation method for eigenvalue distribution functions, we study the temperature dependence of specific heat of the antiferromagnetic Heisenberg model 
on the asymmetric railroad-trestle lattice. This model contains both the sawtooth-lattice and Majumdar-Ghosh models as special cases.
Making extrapolations to the thermodynamic limit using finite size data up to 28 spins, it is found that specific heat of the Majumdar-Ghosh model
has a two-peak structure in its temperature dependence and those of systems near the sawtooth-lattice point have a three-peak structure.
}
\kword{sawtooth lattice, railroad-trestle lattice, Majumdar-Ghosh model, asymmetric leg interaction, 
Heisenberg antiferromagnet, eigenvalue distribution function, multi-peak structure of the specific heat}

\begin{document}
\maketitle
\section{Introduction}\label{sec:1}

Fully frustrated spin systems have been treated extensively in both experimental and theoretical investigations.\cite{Diep} 
By the term of {\it fully frustrated spin system}, we usually mean a system whose classical ground state has infinite continuous degeneracies.\cite{Kubo} 
Some systems have been known to have this property.

A typical example of  fully frustrated spin systems is the antiferromagnetic Heisenberg model (AFHM) on a {\it kagom\'{e}} lattice. 
Harris {\it et al.} studied an ordered ground state of the classical {\it kagom\'{e}} AFHM,\cite{Harris} 
and pointed out that there exist continuous local distortions to the classical ground state without changing the total energy.
The existence of such continuous local distortions leads to the ground state degeneracy $\propto \mbox{const}^N$ with system size $N$.
The quantum dynamics in the spin-1/2 system makes the residual entropy release, and thus the temperature dependence of the specific heat shows a two peak structure.\cite{Sindzingre}
The unique ground state of the quantum system is now believed to be a resonating-valence-bond state.\cite{Mambrini}

Spin-$1/2$ AFHM on a sawtooth lattice (or $\Delta$-chain), which is one of the simplest one-dimensional fully frustrated spin systems,
has been treated by several authors.
Kubo pointed out the twofold-degenerate exact dimer-singlet ground state and a double peak structure
in the specific heat by using the quantum transfer matrix method.\cite{Kubo}
Otsuka applied an approximation method for the eigenvalue distribution function (EvDF) to calculate thermodynamics of AFHM on finite size sawtooth-lattices 
with up to 26 spins under the periodic boundary condition.\cite{Otsuka}
He examined the magnetic-field dependence and estimated lower-temperature peak position of the specific heat in the thermodynamic limit.
The sawtooth lattice is expected to be realized in an oxygen doped Cu based delafossits $\mathrm{YCuO_{2.5}}$\cite{Sen,Simonet} and 
a molecular quantum spin system $\mathrm{Mo_{75}V_{20}}$.\cite{Muller1,Muller2}.
The latter consists of twenty spins, and is often called ``spin doughnut" because of the way of arrangement of $S=1/2$ vanadium ions in the molecular.

A lot of theoretical studies have been done for the spin-$1/2$ antiferromagnetic $XXZ$ model on a railroad-trestle lattice, 
because this model shows interesting quantum phase transitions and critical phenomenena.\cite{Haldane}
Nomura and cowarkers determined the quantitative phase diagram by using the level spectroscopy technique.\cite{Nomura1,Nomura2,Nomura3}
This model contains the Majumdar-Ghosh model as a special case.\cite{Majumdar1,Majumdar2}
The Majumdar-Ghosh model has been known to have the twofold-degenerate exact dimer-singlet ground state, but it is not a fully frustrated system.\cite{Kubo} 

\begin{figure}[b]
\begin{center}
\includegraphics[width=.95\linewidth]{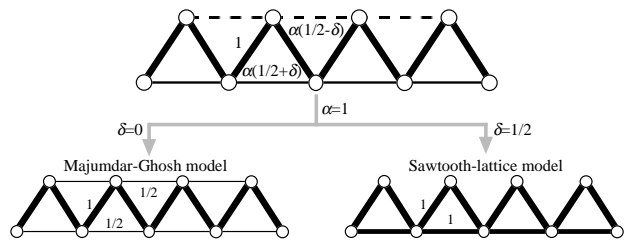}
\end{center}
\caption{Schematic representation of the railroad-trestle lattice with asymmetric leg interactions.
The model reduces to the Majumdar-Ghosh (sawtooth-lattice) model at $\alpha=1$ and $\delta=0$ ($\delta=1/2$).}
\label{fig:1}
\end{figure}

Recently, several authors studied a generalized model, where the strength of the leg interactions in the railroad-trestle lattice model is 
extended to be asymmetric.\cite{Chen1,Sarkar,Capriotti,Nakane} The Hamiltonian is described by
\begin{equation}
     {\cal H} = \sum_{i=1}^N h^{\Delta}_{i,i+1} + \alpha\sum_{i=1}^N \left\{ 1/2 - (-1)^i\delta \right\}h^{\Delta}_{i,i+2},
\label{eq:1}
\end{equation}
with $h^{\Delta}_{i,j} = S^z_iS^z_j+\Delta (S^x_iS^x_j + S^y_iS^y_j)$ (see Fig.~\ref{fig:1}). 
The operator $\mbox{\boldmath{$S$}}_i$ denotes the quantum spin of size $1/2$ on $i$th lattice point, and the periodic boundary condition, i.e., 
$\mbox{\boldmath{$S$}}_{N+1}=\mbox{\boldmath{$S$}}_1$ is imposed.

The parameter $\delta$ measures the asymmetry in the leg interactions.
The ground state phase diagram of this model has been investigated by Nakane, Oguchi and one of the present authors.\cite{Nakane}
Using the level spectroscopy method, they studied how the asymmetry parameter $\delta$ alters phase boundaries.
They found no qualitative effect of $\delta$ on the phase boundaries. In particular, the $\delta$ dependence of the dimer-N\'{e}el phase boundary is quite small.
However, as long as we know, there is no systematic study of the thermodynamic properties of this model.

The main purpose of the present paper is to study the thermodynamic properties of the railroad-trestle lattice model with asymmetric leg interactions,
paying attention to the case of $\alpha=1$. We set $H \equiv {\cal H}_{\alpha=1}$, and concentrate on $H$ below.
It is also assumed that $0\leq \Delta$ and $0\leq \delta \leq 1/2$, and then $H$ has the exact singlet dimer state as the ground state.\cite{Nakane}
It should be noted that $H_{\delta=0}$ is the Majumdar-Ghosh model and $H_{\delta=1/2}$ is the sawtooth-lattice model, as schematically shown in Fig.~\ref{fig:1}. 
As mentioned before, the sawtooth-lattice model with $\delta=1/2$ is a fully frustrated system.\cite{Kubo}  
The thermodynamic properties was already studied, and have been known to have a two peak structure in the temperature dependence of the specific heat.\cite{Kubo,Otsuka} 
The cases of $\delta<1/2$ are not fully frustrated, following the previous definition of this term.
However, the Ising model $H_{\delta,\Delta=0}$ has a rather large residual entropy even for $\delta<1/2$, as described in Sec.~\ref{sec:2} in detail.
This observation motivates us to study the $\delta$ dependence of the specific heat of the Heisenberg model $H_{\delta,\Delta=1}$.

This paper is organized as follows: 
we study the specific heat and entropy of the Ising model $H_{\Delta=0}$ in \S~\ref{sec:2}, in which we apply the transfer
matrix method to this system and the origin of the residual entropy is explained. Our calculation process of thermodynamic properties
by using the EvDF method is explained in \S~\ref{sec:3}. We show the calculated results of the Heisenberg model $H_{\Delta=1}$ in \S~\ref{sec:4},
where we use finite-size data up to 28 sites to make an extrapolation to the thermodynamic limit. 
Finally we summarize our results in \S~\ref{sec:5}.

\section{Thermodynamic Properties at the Ising Limit}\label{sec:2}
In this section, we study the entropy and specific heat of the Ising limit of the our model, $H_{\Delta=0}$. 
In order to use the transfer matrix method, we rewrite the Hamiltonian as the sum of plaquette elements:
\begin{equation}
	H_{\Delta=0}=\sum_{l=1}^{N_{\rm{p}}} h(l),
\label{eq:1}
\end{equation}
where the $l$th plaquette Hamiltonian $h(l)$ is defined by
\begin{eqnarray}
	h(l)&&\hspace{-8mm}= \frac{S^z_{2l-1}S^z_{2l}+S^z_{2l+1}S^z_{2l+2}}{2}+S^z_{2l}S^z_{2l+1}
	\nonumber \\
	&&\hspace{-6mm}
	+\left(\frac{1}{2}+\delta\right)S^z_{2l-1}S^z_{2l+1}+\left(\frac{1}{2}-\delta\right)S^z_{2l}S^z_{2l+2},
\label{eq:1p}
\end{eqnarray}
and $N_{\rm{p}}=N/2$ is the total number of plaquettes (or triangle elements for the sawtooth-lattice case with $\delta=1/2$) 
and assumed a large even number.

The partition function of the system is given by
\begin{eqnarray}
	Z=\sum_{\{S^z\}}\prod_{l=1}^{N_{\rm{p}}} P(S^z_{2l-1},S^z_{2l},S^z_{2l+1},S^z_{2l+2})
\label{eq:2}
\end{eqnarray}
where the sum over $\{S^z\}$ run on all $4^{N_{\rm{p}}}$ spin configurations, and
\begin{equation}
	P(S^z_{2l-1},S^z_{2l},S^z_{2l+1},S^z_{2l+2})=e^{-\beta h(l)}
\label{eq:5}
\end{equation}
with $\beta=1/T$.
It is convenient to introduce a 4$\times$4 transfer matrix $\hat{P}$ by
\begin{eqnarray}
	\hat{P}=
	\left( \begin{array}{cccc}
			e^{-3\beta/4} & e^{-\beta \delta_+/4}
			& e^{\beta\delta_+/4} & e^{\beta/4}\\
			e^{\beta\delta_-/4} & e^{\beta/4}
			& e^{\beta/4} & e^{-\beta\delta_-/4}\\
			e^{-\beta \delta_-/4} & e^{\beta/4} 
			& e^{\beta/4} & e^{\beta\delta_-/4}\\
			e^{\beta/4} & e^{\beta\delta_+/4} 
			& e^{-\beta\delta_+/4} & e^{-3\beta/4}
			\end{array}
		\right),
\label{eq:5}
\end{eqnarray}
where $\delta_{\pm}=1\pm 2\delta$. 
Then, we have
\begin{eqnarray}
	Z=\mbox{Tr}\;\hat{P}^{N_{\rm{p}}}. 
\label{eq:6}
\end{eqnarray}
The largest eigenvalue of $\hat{P}$ is given by
\begin{eqnarray}
	\lambda&&\hspace{-8mm}=\frac{3e^{\beta/4}+e^{-3\beta/4}}{2} 
	\nonumber \\
	&&\hspace{-3mm}
	+\frac{1}{2} \sqrt{5e^{\beta/2}+4e^{\delta\beta}+4e^{-\delta\beta}+2e^{-\beta/2}+e^{-3\beta/2}}.
\end{eqnarray}
Thus we obtain partition function in the limit $N_{\rm{p}} \rightarrow \infty$ as follows:
\begin{eqnarray}
	Z=\lambda^{N_{\rm{p}}},
\label{eq:8}
\end{eqnarray}
which gives the ground state energy per a spin
\begin{equation}
	E(T=0)=\left.-\frac{1}{2}\frac{\partial}{\partial \beta}\log \lambda \; \right|_{\beta=\infty}=-1/8,
\label{eq:8}
\end{equation}
and the residual entropy per a spin
\begin{eqnarray}
	S(T=0)&&\hspace{-8mm}=\left.\frac{1}{2}\left(1-\beta\frac{\partial}{\partial \beta}\right)\log \lambda \; \right|_{\beta=\infty}
	\nonumber \\
	&&\hspace{-8mm}=
	\left\{
	\begin{array}{ll}
	   \frac{1}{2}\log 3 & \mbox{for $\delta=1/2$} \\
	   \frac{1}{2}\log \frac{3+\sqrt{5}}{2} & \mbox{for $0\leq\delta<1/2$} 
	\end{array}
	\right..
\label{eq:8p}
\end{eqnarray}

We here comment on the degeneracy factor of the ground state manifold which gives the energy in eq.~(\ref{eq:8}).
For the sawtooth lattice case with $\delta=1/2$, the degeneracy factor has been already known.\cite{Priour, Fukumoto}
It is easy to see that the ground state energy $E(T=0)=-1/8$ is obtained if each triangle has two antiferromagnetic bonds
and one ferromagnetic bond. There are three ways to choose the position of a ferromagnetic bond in a triangle, 
so that the residual entropy in the first line of eq.~(\ref{eq:8p}) is obtained.

\begin{figure}[t]
\begin{center}
\includegraphics[width=.9\linewidth]{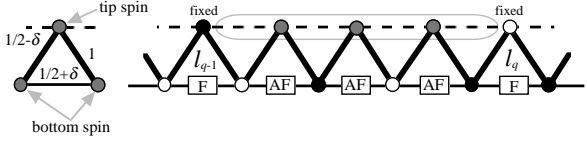}
	\caption{Definition of ``tip spin" and ``bottom spin" (left panel), and a schematic explanation
	of the counting problem of the degeneration factor for a fixed bottom spin configurations (right panel).
	The circles represent spins, and the spin direction of the open circles is opposite to that of the black circles.
	The tip spins represented by open and black circles are fixed tip spins. The exchange coupling surrounded by the gray line
	determines the directions of the gray tip spins between the two  fixed tip spins. }\label{fig:2}
\end{center}
\end{figure}

We turn to the  general case of $\delta\neq 1/2$. It is convenient to introduce the terms, ``tip spin" and ``bottom spin", as shown in Fig.~\ref{fig:2}.
In order to count the degeneracy factor, we first choose a bottom spin configuration, ${\cal S}_{\rm{b}}$.
In ${\cal S}_{\rm{b}}$, some bottom spin pairs are in ferromagnetic state and the others are in antiferromagnetic state.
We assume ${\cal S}_{\rm{b}}$ has $N_{\rm{F}}$ ferromagnetic bottom spin pairs on $l_1$, $l_2$, $\cdots$, $l_{N_{\rm{F}}}$th triangles.
We also define $l_{0}\equiv l_{N_{\rm{F}}}-N_{\rm{p}}$, taking into account the periodic boundary condition.
It should be noted that the direction of the tip spin on a triangle with a ferromagnetic bottom spin pair
is fixed to be opposite to that of the bottom spins in the ground state manifold (see Fig.~\ref{fig:2}).
We now consider how to choose configuration of tip spins between the fixed tip spins on the $l_q$ and $l_{q-1}$th triangles,
which are shown by the gray circles in the right panel in Fig.~\ref{fig:2}.
The tip spin configurations are determined so as the tip spin interaction energy between the two fixed tip spins to be minimized.
The tip spin interaction energy contains $l_q-l_{q-1}$ bonds.
It is easy to observe that the tip spin interaction energy is minimized when we put $l_q-l_{q-1}-1$ bonds
to be antiferromagnetic and the remaining one bond to be ferromagnetic. 
The total number of such tip spin configurations is $l_q-l_{q-1}$, because of the $l_q-l_{q-1}$ ways to choose one ferromagnetic bond.
Therefore, for the fixed bottom spin configuration ${\cal S}_{\rm{b}}$, the degeneracy factor $D({\cal S}_{\rm{b}})$ is given by
\begin{eqnarray}
	 D({\cal S}_{\rm{b}})=\prod_{q=1}^{N_{\rm{F}}}(l_q-l_{q-1}).
\end{eqnarray}
Summing up this degeneracy factor over all bottom spin configurations, then we obtain
\begin{eqnarray}
	e^{NS(T=0)}=\sum_{{\cal S}_{\rm{b}}}D({\cal S}_{\rm{b}})\;\;\;\mbox{for $\delta<1/2$}.
\end{eqnarray}
We have numerically checked that this expression reproduces the second line of eq.~(\ref{eq:8p}).

It is instructive to note that $D({\cal S}_{\rm{b}})$ is altered to be $D({\cal S}_{\rm{b}})=2^{N_{\rm{p}}-N_{\rm{F}}}$ for $\delta=1/2$,
because all configurations of $N_{\rm{p}}-N_{\rm{F}}$ tip spins, other than $N_{\rm{F}}$ fixed tip spins, are possible in the ground state manifold. 
Thus, we obtain the first line of eq.~(\ref{eq:8p}) again as follows:
\begin{eqnarray}
	e^{NS(T=0)}&&\hspace{-8mm}=\sum_{{\cal S}_{\rm{b}}}D({\cal S}_{\rm{b}})
                                                =\sum_{N_{\rm{F}}=0}^{N_{\rm{p}}} \!\;_{N_{\rm{p}}}C_{N_{\rm{F}}} 2^{N_{\rm{p}}-N_{\rm{F}}}
	\nonumber \\
	&&\hspace{-8mm}=3^{N_{\rm{p}}}\;\;\;\mbox{for $\delta=1/2$}.
\end{eqnarray}

\begin{figure}[t]
\begin{center}
	\includegraphics[width=.8\linewidth]{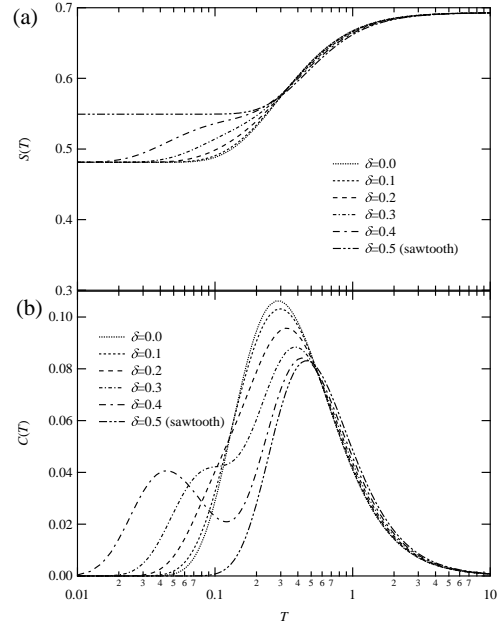}
	\caption{Temperature dependence of (a) entropy $S(T)$ and (b) specific heat $C(T)$ of the Ising limit for
	several values of $\delta$.}\label{Cv:1}
\end{center}
\end{figure}

We show $S(T)$ and $C(T)$ in Fig.~\ref{Cv:1} for several values of $\delta$.
We can confirm in Fig.~\ref{Cv:1} (a) that $S(T)$ decreases toward to $(1/2)\ln 3 \simeq 0.549$ for $\delta=1/2$
and to $(1/2)\ln[ (3+\sqrt{5})/2] \simeq 0.481$ for the others as decreasing temperature.
We also find that $S(T)$ for $\delta=0.4$ and 0.3, which are close to the sawtooth point $\delta=0.5$, are not monotonic:
these have a shoulder around $T\sim 0.1$.
This behavior of the entropy leads to a low temperature peak (or shoulder) in $C(T)$, as seen in Fig.~\ref{Cv:1}(b).

The present results interest us to study effects of the quantum dynamics via spin flip terms.
We are going to study the Heisenberg version of this model in the following sections, where it is shown that
specific heat of the Majumdar-Ghosh model has a two-peak structure in its temperature dependence and those of 
systems near the sawtooth-lattice point have a three-peak structure.

\section{Numerical Method for an Eigenvalue Distribution Function and Thermodynamic Properties}\label{sec:3}

In our calculation of specific heat of the Heisenberg model, we make an estimation of the eigenvalue distribution function (EvDF) with the aid of the Lanczos method
and a sampling technique, as done by Otsuka.\cite{Otsuka} We here give a review of this method, and describe our values of the parameters used in this method.

The EvDF $\rho(\omega)$ is defined as
\begin{eqnarray}
	\rho(\omega)&&\hspace{-8mm}=\mathrm{Tr}\; \delta(\omega-\mathcal{H})\nonumber\\
	                    &&\hspace{-8mm}=-\frac{1}{\pi}\mathrm{Tr} \;\mathrm{Im}\;\mathcal{G}(\omega+i\eta),
	\label{eq:e1}
\end{eqnarray}
where $\mathcal{H}$ is a Hamiltonian, $\mathcal{G}(z)=(z-\mathcal{H})^{-1}$, and $\eta$ is a real positive infinitesimal, which
we approximate as $2\times 10^{-5}$ in our numerical calculation.
Introducing a basis set $\{\phi_n^{(\mu)}|n=1\sim N_{\mu}\}$ in the $\mu$th subspace of $\mathcal{H}$, then the EvDF can be rewritten in the following form:
\begin{eqnarray}
	\rho(\omega)&&\hspace{-8mm}=\sum_{\mu}\rho_{\mu}(\omega)\nonumber\\
	&&\hspace{-8mm}=-\frac{1}{\pi}\sum_{\mu}\sum_{n=1}^{N_{\mu}} \;\mathrm{Im}\;\langle \phi_n^{(\mu)} |\mathcal{G}(\omega+i\eta)|\phi_n^{(\mu)}\rangle.
\end{eqnarray}
The dimension of the subspace, $N_\mu$, is usually so large that we cannot sum over all the basis vectors, which leads us to use a sampling technique. 
In our practical calculation, we samples randomly a set of $N_s$ vectors, $\{\Phi_n^{(\mu)}|n=1 \sim N_s \}$, in such the subspace.
Then $\rho_{\mu}(\omega)$ is approximated by
\begin{eqnarray}
	\rho_{\mu}(\omega)\approx-\frac{c}{\pi}\sum_{n=1}^{N_S}\;\mathrm{Im}\;\langle \Phi_n^{(\mu)} |\mathcal{G}(\omega+i\eta)|\Phi_n^{(\mu)}\rangle,
\end{eqnarray}
with the normalization constant $c$, which is determined from the condition $\int \rho_{\mu}(\omega)d\omega=N_{\mu}$.
This type of sampling was used by Imada and Takahashi in their quantum transfer Monte Carlo method,\cite{Imada}
where it was discussed that enlargement of system size gets $N_S$ to be reduced to evaluate physical quantities within a fixed accuracy.

We turn to the estimation of $\langle \Phi_n^{(\mu)} |\mathcal{G}(z)|\Phi_n^{(\mu)}\rangle\equiv \mathcal{G}_{n,n}^{\mu}(z)$.
We make a tri-diagonalization of $\mathcal{H}$ by the Lanczos method, setting $|\Phi_n^{(\mu)}\rangle$ as the initial vector.
Using the resultant Lanczos coefficients $\{\alpha_l,\beta_l|l=1,2,\cdots\}$,
the matrix element $\mathcal{G}_{n,n}^{\mu}$ can be expressed as the following continued fraction form:
\begin{eqnarray}
	\mathcal{G}_{n,n}^{\mu}(z)=\frac{1}{\displaystyle z-\alpha_1-\frac{|\beta_1|^2}{\displaystyle z-\alpha_2-\frac{|\beta_2|^2}{\ddots}}}.
\end{eqnarray}
It has been known that the convergence of the continued fraction form is so fast that we only need to calculate first 50-200 Lanczos coefficients.\cite{Otsuka}
Our typical iteration number is about 100, for which we checked the convergence of calculated quantities using 16-site systems.

In order to estimate thermodynamic properties, we have to calculate the $m$th moment
\begin{eqnarray}
	F_m(\beta)=\int_{-\infty}^{\infty}\omega^m\rho(\omega)e^{-\beta \omega}d\omega
\end{eqnarray}
as a function of $\beta$. It is not efficient to carry out the numerical integration over $\omega$ at each value of $\beta$,
because $\rho(\omega)$ fluctuates much greater than the other factors.
Taking into account this fact, we first calculate the histogram
\begin{eqnarray}
	\tilde{\rho}(\omega_i)=\int_{\omega_i-\delta\omega/2}^{\omega_i+\delta\omega/2}\rho(\omega)d\omega,
\end{eqnarray}
with the energy mesh $\delta\omega$. This numerical integration should be carried out carefully,
because $\rho(\omega)$ consists of Lorentzians with a very small width of $\eta$.
If we choose the energy mesh as $\beta\delta\omega/2 \ll 1$, then
\begin{eqnarray}
	e^{-\beta(\omega_i\pm\delta\omega/2)}\approx e^{-\beta\omega_i}\left(1\mp\frac{\beta\delta\omega}{2}\right)\approx e^{-\beta\omega_i}
\end{eqnarray}
holds with high accuracy. Therefore, the calculation of $m$th moment reduces to
\begin{eqnarray}
	F_m(\beta)=\sum_i \omega_i^m \tilde{\rho}(\omega_i) e^{-\beta \omega_i},
\end{eqnarray}
which give us much efficient way to evaluate the temperature dependence.
Once the temperature dependence of the moments is obtained, the specific heat per a spin, $C(T)$, is calculated by
\begin{eqnarray}
	C(T)=\frac{\beta^2}{N} \left\{\frac{F_2(\beta)}{F_0(\beta)}-\left[\frac{F_1(\beta)}{F_0(\beta)}\right]^2\right\}.
\end{eqnarray}
In our calculation, setting our maximum inverse temperature as $\beta_{\rm{max}}=100$, we adopt $\delta\omega=2\times10^{-4}$
so as to be $\beta\delta\omega/2\leq 10^{-2}$. 

Finally, we introduce our extrapolation method from finite-size data to the thermodynamic limit $N\to\infty$.
We employ the next equation as the extrapolation function:
\begin{eqnarray}
	C(T,N)=C(T,N\to\infty)+\frac{A}{N^{\kappa}}\hspace{0.3cm} (A:\mbox{Constant}) \label{eq:19}
\end{eqnarray}
where $\kappa$ denotes a positive integer number and later we consider about $\kappa$.

\section{Multi-Peak Structure of the Specific Heat of AFHM on the Railroad-Trestle Lattice
 with Asymmetric Leg Interactions}\label{sec:4}

In this section, we present our numerical results at the Heisenberg point $\Delta=1$ based on the EvDF method.
The Hamiltonian treated here is
\begin{eqnarray}
	H_{\Delta=1}&&\hspace{-8mm}=\sum_{i=1}^N\mbox{\boldmath{$S$}}_i\cdot\mbox{\boldmath{$S$}}_{i+1}+\sum_{i=1}^{N/2}\left[ \left(\frac{1}{2}-\delta \right)
	\mbox{\boldmath{$S$}}_{2i}\cdot\mbox{\boldmath{$S$}}_{2(i+1)}\right.
	\nonumber \\
	&&\hspace{-8mm}
	\left.+\left(\frac{1}{2}+\delta \right)\mbox{\boldmath{$S$}}_{2i-1}\cdot
	\mbox{\boldmath{$S$}}_{2i+1}\right].\label{eq:h1}
\end{eqnarray}
Before describing results for $0 \leq \delta < 0.5$, which is the main concern in this paper,
we examine our calculations at $\delta = 0.5$, where we can use the existing results.\cite{Otsuka}

\subsection{Sawtooth-Lattice Point $\delta=0.5$}\label{sub:1}

%
\begin{figure}[b]
\begin{center}
	\includegraphics[width=.8\linewidth]{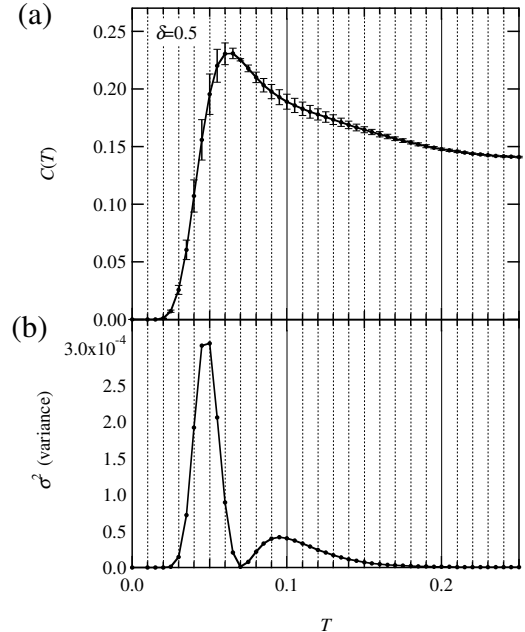}
	\caption{Calculated results of (a) $C(T)$ for the sawtooth-lattice AFHM with $N=22$,  and (b) the unbiased variance $\sigma^2(T)$.}\label{var:1}
\end{center}
\end{figure}

In Fig.~\ref{var:1}, we show low-temperature specific heat of the 22-site sawtooth-lattice AFHM together with the unbiased variance $\sigma^2$.
Here, $N_s=250$ is chosen to calculate $C(T)$, and we divide the 250 samples into 5 sets of 50 samples to get the variance, i.e.,
\begin{eqnarray}
	\sigma^2(T)=\frac{1}{4}\sum_{j=1}^5[C(T)-C_j(T)]^2, \label{}
\end{eqnarray}
where $C_j(T)$ represents the specific heat calculated with the $j$th set.
We find in Fig.~\ref{var:1} that the sampling error becomes severe when temperature decreases and $C(T)$ varies rapidly.
As described later, sampling errors at temperatures below the peak temperature, $T_p \simeq 0.06$, poses an obstacle to our extrapolation procedure
at low temperatures.

\begin{figure}[b]
\begin{center}
	\includegraphics[width=.8\linewidth]{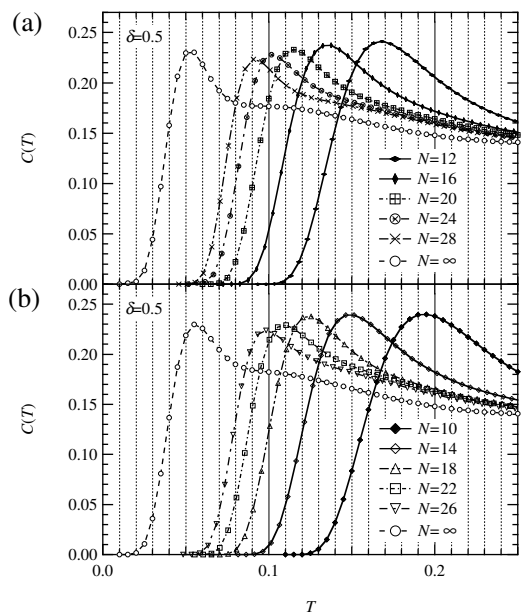}
	\caption{Finite-size and extrapolated results of $C(T)$ for the sawtooth-lattice AFHM, where $N/2$ is even in (a) and odd in (b).
	The offset in the horizontal axis by $1/N$ is applied for the $N$-spin results.
	}\label{Cv:5}
\end{center}
\end{figure}
\begin{figure}[b]
\begin{center}
	\includegraphics[width=.85\linewidth]{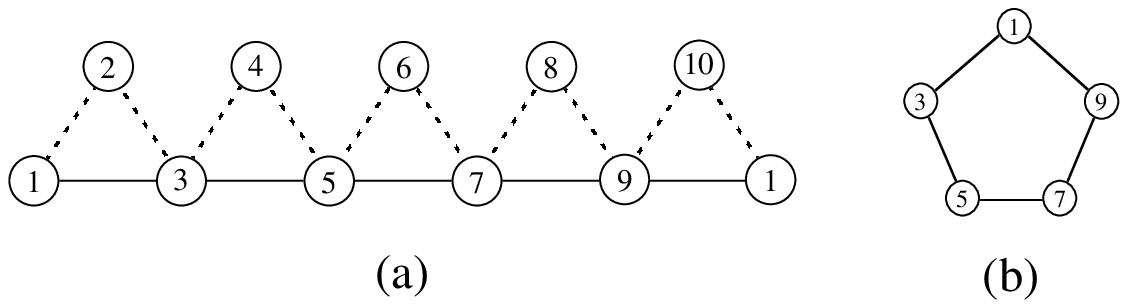}
	\caption{Graphical representations of (a) 10-site sawtooth lattice and (b) its bottom structure.
	}\label{fig:3}
\end{center}
\end{figure}

We now study the dependence of $C(T)$ on the system size $N$ (even number).
We show $C(T)$ in Fig.~\ref{Cv:5}, classifying by $N/2$ being an even or odd number to obtain smooth evolution on the system size.
If we don't make such a classification, for example, the peak height as a function of $N$ shows a nonmonotonic behavior.
In order to understand this behavior, we focus on the bottom structure of $N$-site sawtooth lattice. 
A $N$-site sawtooth lattice has a bottom shaped polygon which has $N/2$ angles.
An example is given in Fig.~\ref{fig:3}, where we show a sawtooth lattice with $N=10$ whose bottom structure is a pentagon.
Let us assume that the coupling for the bonds represented by the dashed line in Fig.~\ref{fig:3} can be omitted.
Then, only the bottom polygon contributes to specific heat.
The ground-state of the pentagon in Fig.~\ref{fig:3}(b) consists of one ferromagnetic bond and four antiferromagnetic bonds.
We call this situation ``$1/5$ frustration".
On the other hand, if the bottom structure is square, hexagon and so on, then there exists no frustration. 
Thus, if $N/2$ is odd number, the bottom polygon in the $N$-site sawtooth system has $2/N$ frustration.
In this argument, we have omitted the dashed-line bonds. 
However, the nonmonotonic behavior on $N$ in our calculated data seems to indicate that such a simple picture holds even when the omitted couplings are again taken into account.
In our extrapolation procedure, the two data sets are treated independently and the consistency between the two extrapolated results is used to check our procedure.

\begin{figure}[h]
\begin{center}
	\includegraphics[width=.8\linewidth]{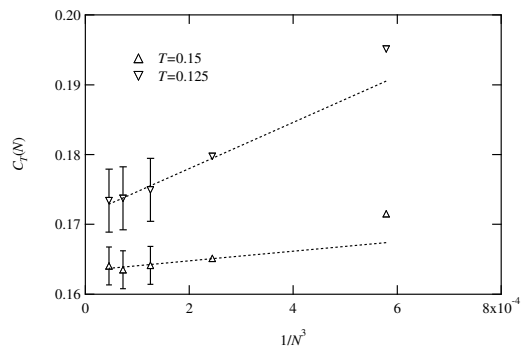}
	\caption{Plot of $C_T(N)$ vs $1/N^3$ at $T=0.15$ and 0.125, where $\delta=0.5$. 
	}\label{size:1}
\end{center}
\end{figure}

We turn to the index $\kappa$ defined in (\ref{eq:19}).
To estimate an adequate value of $\kappa$, we examine the specific heat as a function of $N$ for fixed temperatures, $T=0.15$ and $0.125$,
at which sampling error is not so large that the size dependence is not smeared.
In Fig.~\ref{size:1} the specific heat $C_T(N)$ against $1/N^3$ is shown.
We find a linear behavior for the large systems, and thus we employ the extrapolation function (\ref{eq:19}) with $\kappa=3$ in the case of sawtooth lattice.
The extrapolation results for $N\rightarrow\infty$ are shown in Fig.~\ref{Cv:5} as the open circles.
The extrapolation by the even-$N/2$ data set in Fig.~\ref{Cv:5}(a) is consistent with that by the odd-$N/2$ one, as expected.

Otsuka calculated the specific heat by the same method for odd-$N/2$ systems up to $N=26$.\cite{Otsuka}
Our calculated results are in good agreement with his results at temperatures higher than the peak temperature $T_p$.
Although both results are also consistent with each other within the sampling error at $T<T_p$,
there exists small deviation in average values. Magnitude of the deviation is order of the marker size in Fig.~\ref{Cv:5}.
This small deviation gets our extrapolation result to be somewhat larger than Otsuka's extrapolation at low temperatures.
Expanding the sampling number may resolve this problem. 
However, we don't pursue this problem, because it is time-consuming and does not change our qualitative conclusions.

\subsection{Majumdar-Ghosh Point $\delta=0$}\label{sub:2}

We here study the Majumdar-Ghosh model with $\delta=0$.
This model is well known for the exact singlet-dimer ground state.\cite{Majumdar1,Majumdar2}
However, as long as the present authors know, the thermodynamic property has never been examined.

\begin{figure}[b]
\begin{center}
	\includegraphics[width=.8\linewidth]{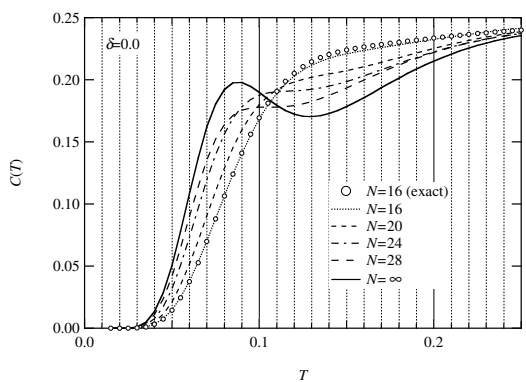}
	\caption{Low-temperature specific heat of the Majumdar-Ghosh model with $\delta=0.0$
	for finite-size systems and the thermodynamic limit.
	The open circles represent the exact specific heat for the 16-site system.
	}\label{Cv:2}
\end{center}
\end{figure}

We show the temperature dependence of the low temperature specific heat for finite size systems with $N=16$, 20, 24 and 28 in Fig.~\ref{Cv:2}.
(For the $N=16$ system, the exact result is also shown as the open circles, which are in good agreement with the EvDF result.)
We find in Fig.~\ref{Cv:2} that a shoulder appears around $T=0.1$ as the system size is enlarged.
This size dependence make us anticipate the appearance of a low-temperature peak in the thermodynamic limit.
In fact, following the extrapolation procedure explained in the previous section, we obtain a low-temperature peak in our extrapolation shown by the solid line in Fig.~\ref{Cv:2}.
The peak temperature is estimated as $T^{\delta=0}_p \approx 0.09$, which is higher than that of the sawtooth-lattice model, $T^{\delta=1/2}_p \approx 0.06$.

\begin{figure}[b]
\begin{center}
	\includegraphics[width=.8\linewidth]{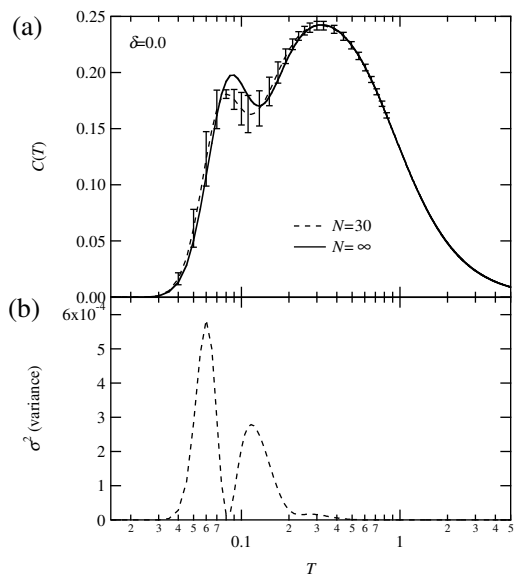}
	\caption{(a) Specific heat for the whole temperature range for the Majumdar-Ghosh model with $N\rightarrow\infty$ and our preliminary result of 
	the 30 site system. (b) Unbiased variance $\sigma^2$ for the $N=30$ result.
	}\label{Cv:30}
\end{center}
\end{figure}

In Fig.~\ref{Cv:30}, we show the specific heat of $N\rightarrow\infty$ for a wider temperature range.
It is found that a high-temperature peak exists at $T\approx 0.3$, which is almost same as the peak temperature of the Ising limit (see Fig.~3).
This result indicates that the spin flip terms in the Majumdar-Ghosh model get the residual entropy of Ising limit to release and lead to the low-temperature peak at $T\approx 0.09$.

The two-peak structure of the specific heat for the Majumdar-Ghosh model is one of the main conclusions in this paper.
In order to check this further, we carry out a preliminary calculation on the 30-site Majumdar-Ghosh model.
The result is shown in Fig.~\ref{Cv:30} as the dashed line, together with sampling error.
Although it has large sampling error due to smallness of the sampling number, 
it suggests that the 30-site Majumdar-Ghosh model has a two-peak structure in the temperature dependence of the specific heat,
which gives a support to our conclusion.

\subsection{Evolution of the Specific Heat as a Function of $\delta$}\label{sub:3}

\begin{figure}[b]
\begin{center}
	\includegraphics[width=.8\linewidth]{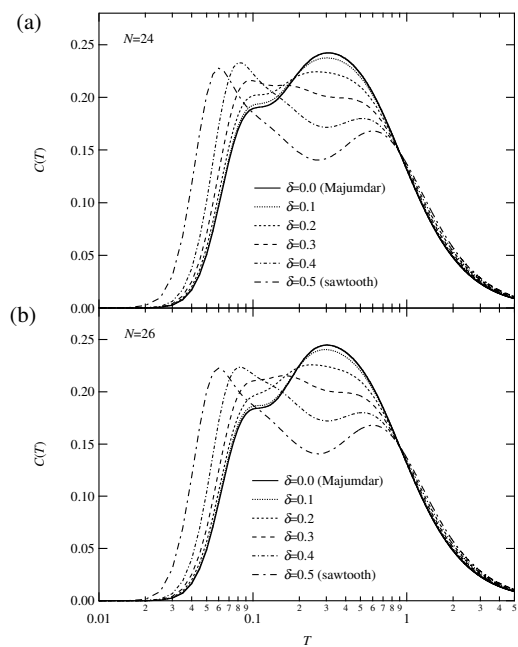}
	\caption{Overall behavior of $C(T)$ for the AFHM on the asymmetric railroad-trestle lattice with (a) $N=24$ and (b) $N=26$
	for selected values of $\delta$.
	}\label{Cv:size}
\end{center}
\end{figure}

Finally, we study how $C(T)$ depends on the asymmetry parameter $\delta$.
In Fig.~\ref{Cv:size}, we show $C(T)$ of the 24- and 26-site systems for $\delta=0.0,\;0.1,\cdots,\;0.5$.
In both systems, $C(T)$  for $\delta=0.0,\;0.1$ and 0.2 has a high-temperature peak at $T\simeq 0.3$ and a shoulder around $T=0.1$.
For $\delta=0.3$, the high-temperature peak and the shoulder merge into a terracing.
For $\delta=0.4$ and 0.5, $C(T)$ shows a two-peak structure clearly.
In addition, a small shoulder at $T\simeq 0.1$ appears.

\begin{figure}[t]
\begin{center}
	\includegraphics[width=.8\linewidth]{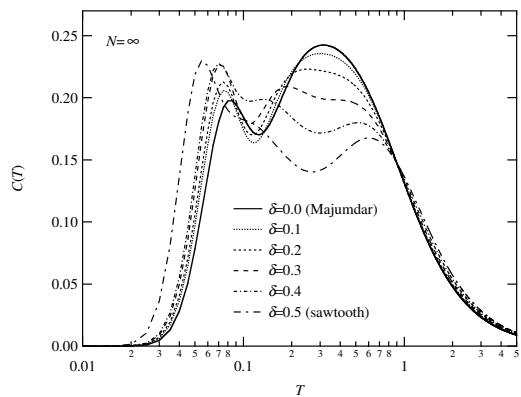}
	\caption{Estimations of $C(T)$ in the thermodynamic limit for the AFHM on the asymmetric railroad-trestle lattice.
	}\label{Cv:6}
\end{center}
\end{figure}

Making extrapolation to the thermodynamic limit, we obtain the results shown in Fig.~\ref{Cv:6}. 
Comparing with the finite size data, we find the low-temperature shoulder for $\delta=0.0,\;0.1$ and 0.2 changes to a low-temperature peak.
Also, the low-temperature side edge of the terracing for $\delta=0.3$ is separated from it and turns into a low-temperature peak.
For $\delta=0.4$, the small shoulder at $T\simeq 0.1$ changes to a small peak, and thus $C(T)$ for $\delta=0.4$ has three peaks.
The small shoulder for $\delta=0.5$ becomes clearer, but it does not turn to a peak. 
Thus, we conclude that the present model has a two-peak structure in the specific heat,
and an additional mid-temperature peak appears for $\delta\simeq 0.4$.

To understand the overall behavior of the specific heat, it is helpful to make a comparison with the result of the Ising limit.
As for $\delta$-dependence of the high-temperature peak, the Heisenberg system has similar features to the Ising limit as seen in Figs. 3 and \ref{Cv:6}:
for the both cases, the high-temperature peak position as a function of $\delta$ is almost same, and the peak height is decreased as $\delta$ increasing.
The lowest-temperature peak in Fig.~\ref{Cv:6}, which originates in the spin flip terms, grows when $\delta$ increases.
This observation can be interpreted in terms of the Ising limit: we find in Fig.~3(a) that, at low temperatures, the entropy is larger for larger $\delta$,
and thus the spin flip term can potentially lead to larger lowest-temperature peak for larger $\delta$.
Also, in terms of the Ising limit, the three-peak structure for $\delta=0.4$ in Fig.~\ref{Cv:6} is plausible, because the specific heat has already two peaks in the Ising limit.
Finally, we notice in Fig.~\ref{Cv:6} that $C(T)$ becomes to be more sensitive to $\delta$ as $\delta$ increasing.
We can find the same tendency in Fig.~3 for the Ising limit.

%

\section{Summary}\label{sec:5}

In this paper, we consider thermodynamic properties of the Ising and Heisenberg antiferromagnets on a railroad-trestle lattice with asymmetric leg interactions. 
At first we have investigated entropy structure of the sawtooth-lattice and railroad-trestle-lattice Ising antiferromagnets,
where we have found that the railroad-trestle-lattice model has less residual entropy than the sawtooth-lattice model 
because of difference in total number of possible tip-spin configurations in the ground state manifold. 
Next we have calculated thermodynamics of the Heisenberg antiferromagnets by using an eigenvalue distribution function. 
Our obtained results are summarized as follows.
(1) We have found a finite-size effect: we can classify finite-size results of $C(T)$ according to $N/2$ being even or odd. 
This tendency becomes to be more pronounced for systems with larger asymmetry.
(2) In the thermodynamic limit, our calculated results indicate that the Heisenberg antiferromagnet on the asymmetric railroad-trestle lattice has two or more peaks
in the temperature dependence of the specific heat. Especially, the specific heat has a three peak structure for $\delta \simeq 0.4$, and a two peak structure even at $\delta=0$. 
For $\delta=0$, the Majumdar-Ghosh model, we estimate the lower-temperature peak position as $T_p\approx 0.09$.
We have also pointed out that the $\delta$ dependence of specific heat of the Heisenberg model is possible to be interpreted in terms of the Ising limit.

As for finite size data themselves, we have not observed multi-peak structure in $C(T)$ at $\delta<0.3$ within the present calculations for $N \leq 28$,
though the size dependence of $C(T)$ exhibits a sign of the multi-peak structure. 
As a future problem, it is interesting to study the specific heat of larger systems than $N=28$ by using the density matrix renormalization group method,\cite{Shibata}
which provides a critical check of the conclusion obtained in this paper.

\acknowledgements
We have used a part of the code provided by H. Nishimori in TITPACK Ver.2.

\end{document}